\documentstyle[epsfigx]{mn}
\newcommand{\HII}{H\,{\sc ii}}
\newcommand{\ub}{U_{\rm B}}
\newcommand{\urad}{U_{\rm rad}}
\newcommand{\diff}{\rm d}

\newcommand{\psyn}{P_{syn}}
\newcommand{\ptherm}{P_{therm}}
\newcommand{\prad}{P_{rad}}
\newcommand{\pnu}{P_{rad,1.5~{\rm GHz}}}
\newcommand{\lfir}{L_{FIR}}
\newcommand{\ms}{M_\odot}
\newcommand{\rvek}{{\bf r}}
\newcommand{\rx}{R_{x,i}}
\newcommand{\ry}{R_{y,i}}
\newcommand{\rz}{R_{z,i}}
\newcommand{\apjs}{ApJS }
\newcommand{\apj}{ApJ }
\newcommand{\aajou}{A\&A }

\newcommand{\mnras}{MNRAS }

\newcommand{\aj}{AJ }
\begin{document}
\title{Constraints on Cosmic Ray propagation from Radio
Continuum data of NGC~2146}
\author[U.~Lisenfeld et al.]
{U.~Lisenfeld\thanks{present address: Osservatorio Astrofisico die Arcetri,
Largo E. Fermi 5, 50125 Florence, Italy, e-mail: ute@arcetri.astro.it}
, P.~Alexander, G.G.~Pooley, T.~Wilding\\
 Mullard Radio Astronomy Observatory,
        Cavendish Laboratory,
        Madingley Road,
        Cambridge CB3 OHE }
\maketitle
\begin{abstract}
We present high-sensitivity multi-frequency
radio continuum observations of the starburst galaxy NGC~2146.
We have fitted these data with a three-dimensional diffusion model.
The model can describe the radio emission from the inner disk of
NGC~2146 well, indicating that diffusion is the dominant mode
of propagation in this region.
Our results are indicating that NGC~2146 has recently
undergone a starburst, the star forming activity being located in a 
central bar.
The spatial variation of the radio emission and of the spectral
index yield
tight constraints on the diffusion coefficient $D_0$ and the
energy dependence of the diffusion.
Away from the central bar of the galaxy the radio emission
becomes filamentary and the diffusion model was found to be a poor fit to the
data in these regions; we attribute this to
different transport processes
being important in the halo of the galaxy.
\end{abstract}

\section{Introduction}
Multifrequency radio observations of nearby,
face-on galaxies present a unique way to study the
Cosmic Ray (CR) propagation within their disks.
These studies provide constraints on the main parameters of this propagation,
i.e. the value and energy dependence of
the diffusion coefficient and the form of the electron energy distribution
immediately after acceleration.  Furthermore,
because of the relatively long radiative life-time of CR electrons
(which is in normal, non-starburst galaxies
up to several $10^7$ years) such observations also provide
information about the temporal evolution of
the sources of the CR's on the same time scale; in turn, since the sites
of acceleration are supernova remnants the data are probing the
history of star formation on time scales up to of order $ \approx 10^8$ years.

To date, the parameters describing
CR propagation have only been derived for our Galaxy;
radio observations are the only way of obtaining
information for external galaxies.
In this paper we analyse high-sensitivity multi-frequency radio data for
the nearby galaxy NGC~2146.  Our data have sufficient frequency coverage
that we can include in our model fits both a thermal and diffuse synchrotron
component for the radio emission.

NGC~2146 is a very luminous radio source ($\pnu=5.9\,10^{22}$~W/Hz)
at a distance of 21.8 Mpc (Benvenuti, Capacioli \& D'Odorico 1975, adjusted to
$H_0$=50 km s$^{-1}$ Mpc). Most of its radio luminosity
is due to a compact starburst in its centre.
The extent of the radio emission
(1.5 arcmin $\approx$ 9.6 kpc) is much less than the extent of the
optical emission ($\approx$ 6 arcmin). This suggests that active star
formation (both current and in the recent past) is confined to
this central region of NGC~2146. In this paper we are only discussing
this inner region of the galaxy ($\approx$ 1.5 arcmin $\times$ 0.7 arcmin) that is 
characterized by the 
strong radio emission. 
Optical images of NGC~2146 show it
to be a peculiar spiral galaxy (e.g. Benvenuti et al. 1975).
Hutchings et al. (1990) have presented
a detailed study of NGC~2146 in several wavebands and proposed that
the galaxy is in the late stages of a merger which has given rise
to an intense burst of star formation in the nucleus.

\section{Observations}
\begin{figure*}
\begin{minipage}{150mm}
\centerline{
\epsfigx{file=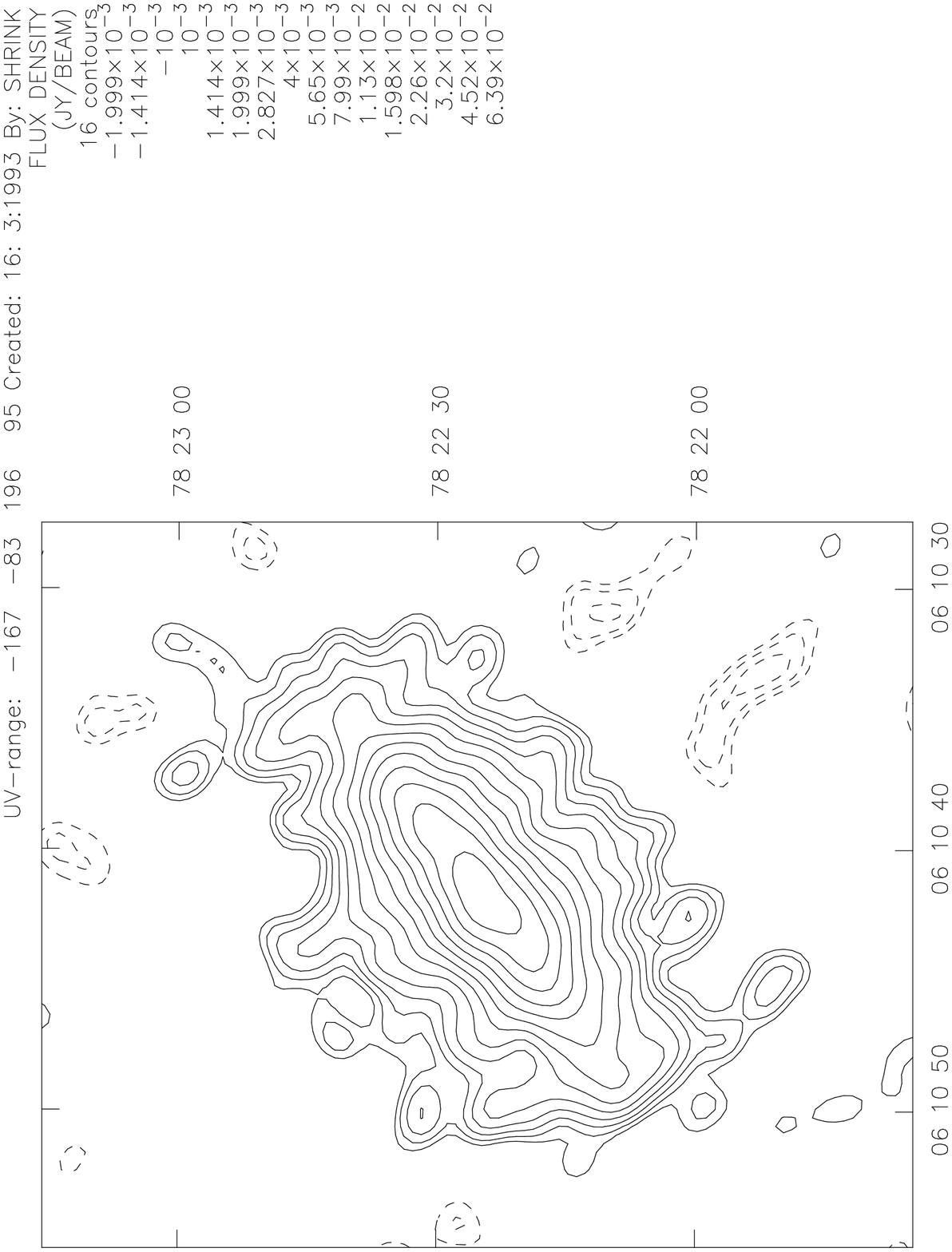,width=6.5cm,clip=,angle=270}{2}\quad
\epsfigx{file=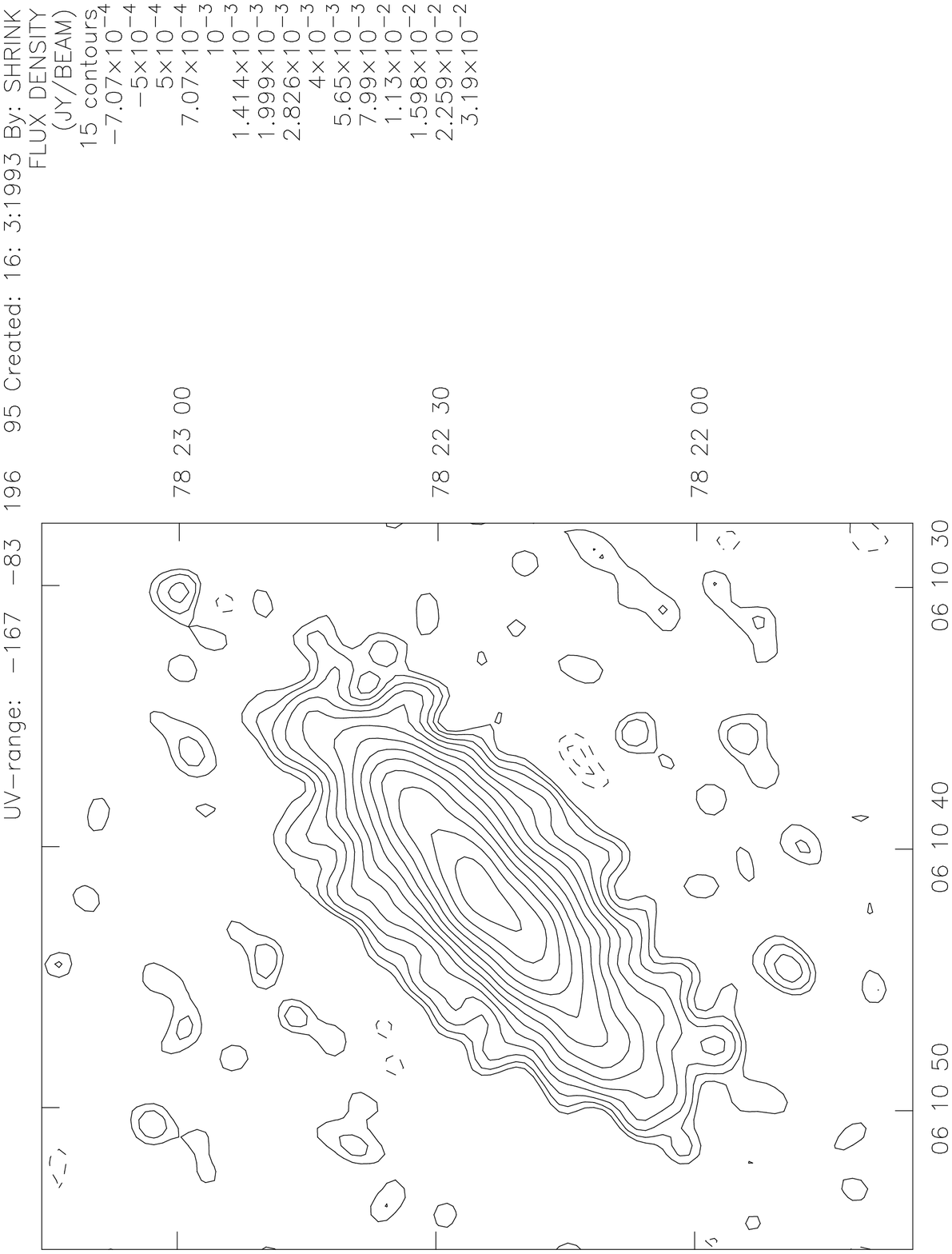,width=6.5cm,clip=,angle=270}{2}}
\smallskip
\centerline{
\epsfigx{file=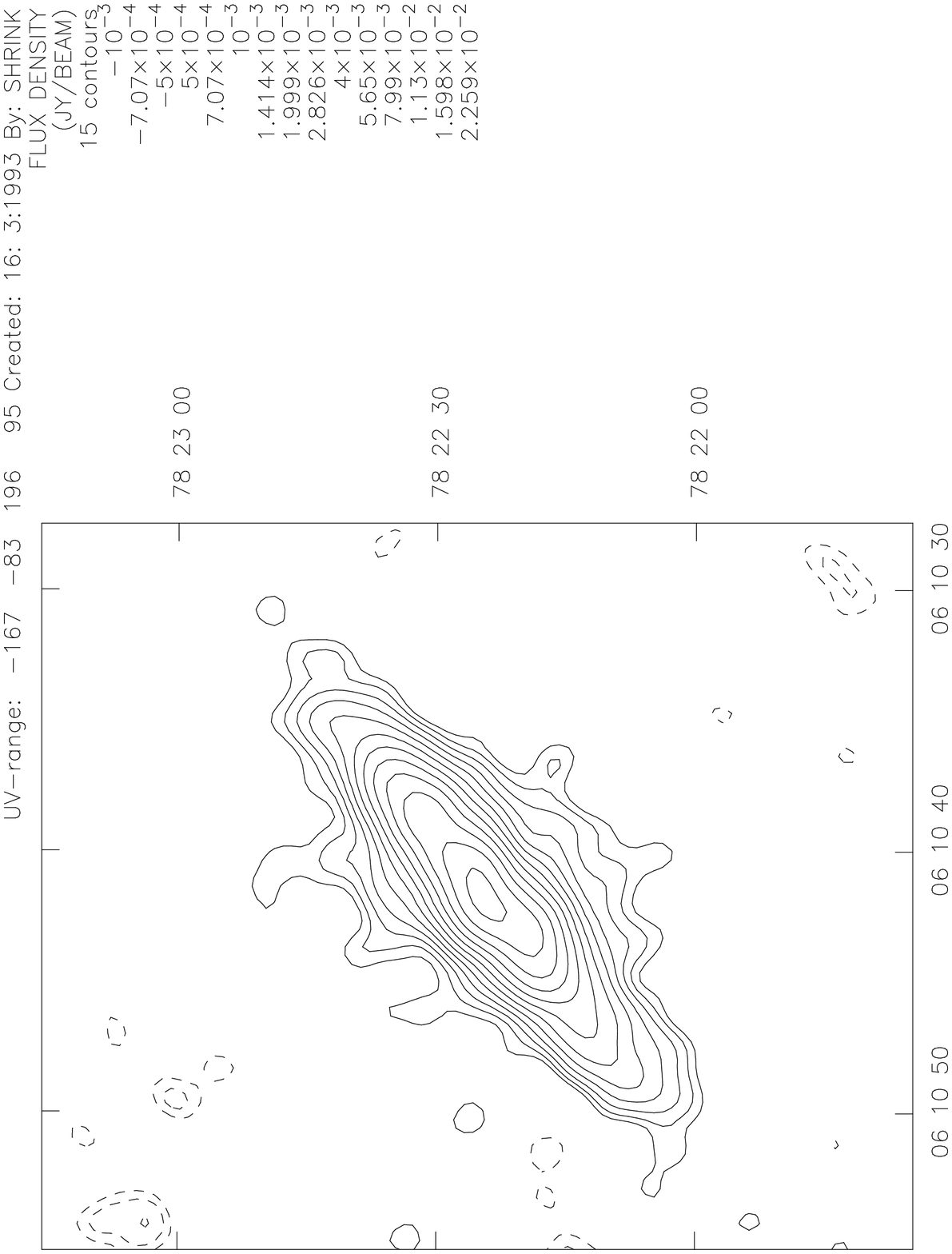,width=6.5cm,clip=,angle=270}{2}\quad
\epsfigx{file=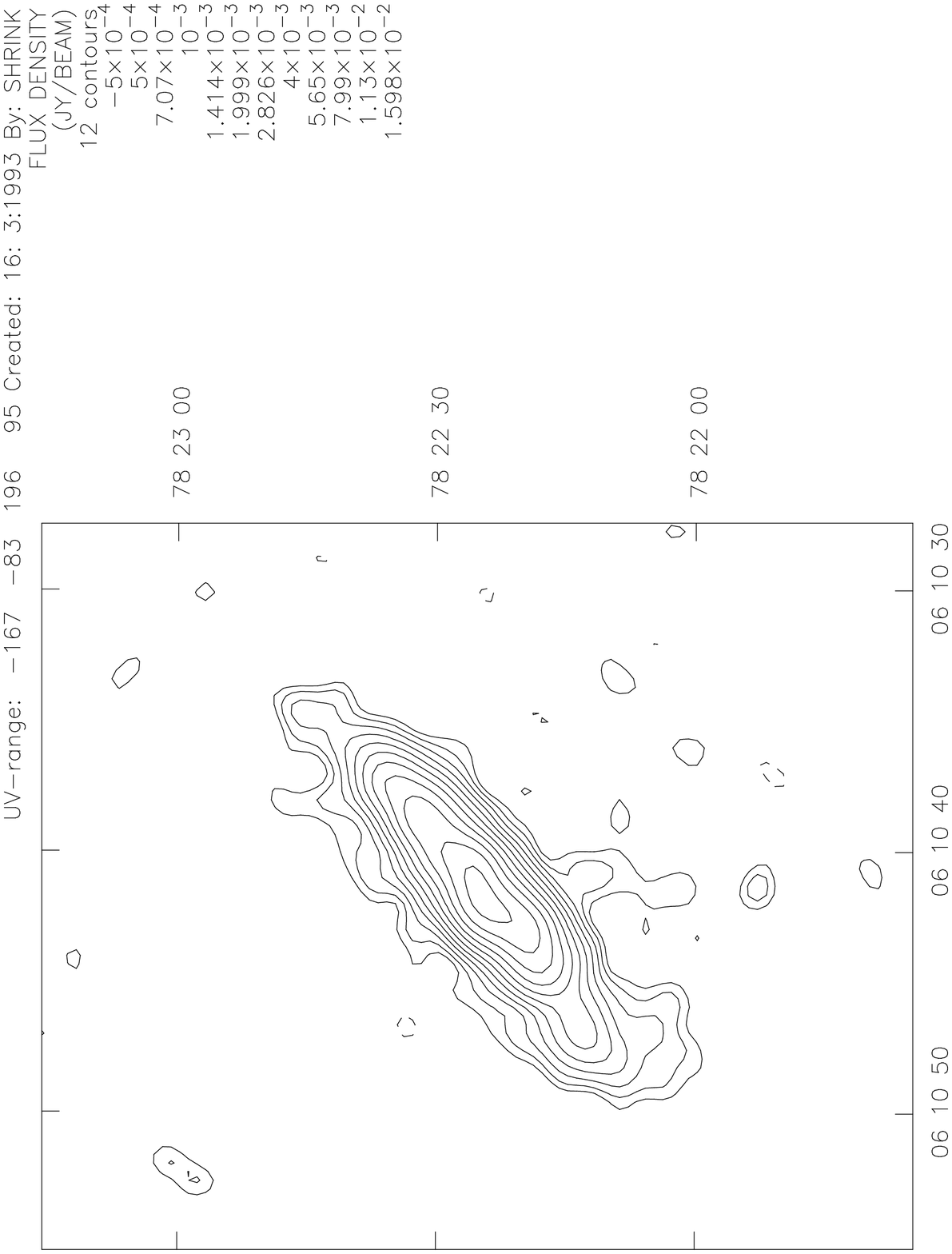,width=6.5cm,clip=,angle=270}{2}}
\caption[Maps of NGC~2146]{Maps of NGC~2146 at 1.5, 5, 8.4 and 15 GHz at
a resolution of
4.7 $\times$ 4.7 arcsec$^2$. The contour levels start at 1 mJy/beam
(1.5GHz) and  0.5 mJy/beam (for the other frequencies),
and increase by factors of $\sqrt 2$. The r.m.s. noise levels are:
0.67 mJy/beam (1.5 GHz), 0.244  mJy/beam (5 GHz),
0.18 mJy/beam (8.4 GHz) and
0.16 mJy/beam (15 GHz).}
\end{minipage}
\end{figure*}

NGC~2146
was observed with the VLA at 8.4 GHz and 327~MHz, and
with the Cambridge 5-km Ryle Telescope (RT) at 5 and 15 GHz.
Additionally, existing VLA data at 1.5~GHz (Condon 1983) were available.
When the observations of NGC~2146 were made, the RT was operated with
a band-width of 280~MHz split into 28 10-MHz frequency channels;
together with the minimum baseline of 18~m this provides excellent
temperature sensitivity  at both 5 and 15~GHz.

For the RT observations a total of five telescope configurations at each
observing frequency were employed giving a nearly fully filled aperture
out to the maximum baseline.  A minimum of two 12-hour runs in each
configuration were obtained; any data of poor quality were rejected.
This gave a resolution approximately
$1.0 \times 1.0\mbox{\rm cosec}( \delta )$~arcsec$^{2}$ at 15 GHz and
$3.0 \times 3.0\mbox{\rm cosec}( \delta )$~arcsec$^{2}$ at 5 GHz.
At 5~GHz observations of phase-calibrators were made at the beginning
and end of each run, while at 15~GHz calibration observations were
interleaved with those of NGC~2146.  3C~286 and 3C~48 were observed
regularly as flux calibrators.
Calibration and data-editing (principally to remove narrow-band and
time-varying interference)
was performed in the MRAO package {\em Postmortem}, with subsequent reduction
in AIPS and the MRAO package {\em Anmap}.

The 8.4-GHz VLA observations were in the B, C and D arrays with the
327~MHz observations in A array only; the observations were
performed during 1990/91.  Reduction of the data followed standard VLA
procedures with calibration, editing and imaging performed in AIPS.

A number of maps were made at each frequency by tapering the aperture-plane
to highlight structure on different angular scales.
For the spectral comparison we used only the four highest frequencies and
all the maps were made with very similar aperture plane coverage,
CLEANed and restored (or convolved in the case of the 1.49~GHz image)
to a common resolution of
$4.7 \times 4.7$~arcsec$^{2}$.
All maps have been corrected for the primary beam response of the
telescope used to make the observations.
This set of four images is shown in Fig. 1, together with a high-resolution
image at 15~GHz in Fig. 2 and the 327~MHz image in Fig. 3.
In Fig. 4 a map of the spectral index between 1.5 and 8.4 GHz is shown.

\section{Radio properties of NGC~2146}

Most of the radio emission of NGC~2146 lies along an elongated, 
bar-like structure in position angle
$143\pm 3^\circ$ coincident with the optical major axis
(143$^\circ$, Benvenuti et al. 1975). This elongated shape could
be a disk viewed from a high inclination or it could be due 
to a real bar-like shape of the sources of the CR's.
From the radio data at one frequency alone it is not possible to
distinguish between these two cases. However, as will be shown
in the following, the variations of the radio 
spectral index  together with a model for the CR
propagtion allow to draw conclusions about the shape
of the CR sources.
In the high resolution image
(Fig. 2), the sources at 15 GHz show a distinctive {\bf S} shape that
had already been noticed in earlier observations by
Kronberg $\&$ Biermann (1981).
The central region is shown in our highest resolution images to
contain a number
of luminous unresolved sources superimposed on the
general background emission.

At higher frequencies NGC~2146 appears more compact and shows
relatively little emission away from the major axis of the radio
emission while
at lower frequencies the emission has an irregular appearance
and several extensions perpendicular to the
major axis which are prominent particularly
in the 1.5~GHz and the 327~MHz images.

\begin{figure}
\centerline{
\epsfigx{file=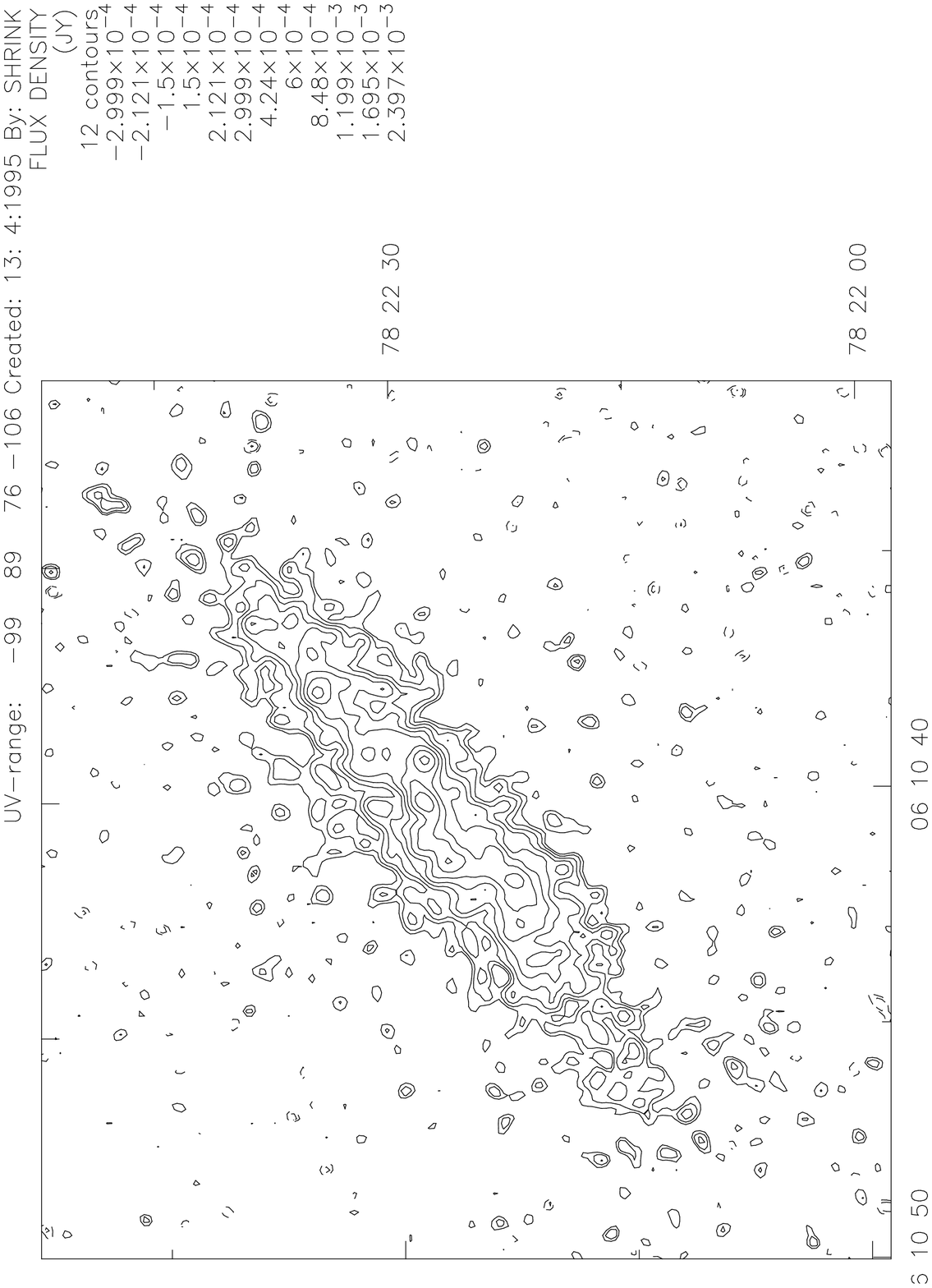,width=6.5cm,clip=,angle=270}{2}}
\caption[Map at 15 GHz]{
Map of NGC~2146 at 15~GHz at a resolution of 1.18 $\times$ 1.05 arcsec$^2$.
The contour levels start at 0.15 mJy/beam, 
and increase by factors of $\sqrt 2$. 
The r.m.s. noise level is at 0.076 mJy/beam.
}
\end{figure}
\begin{figure}
\centerline{
\epsfigx{file=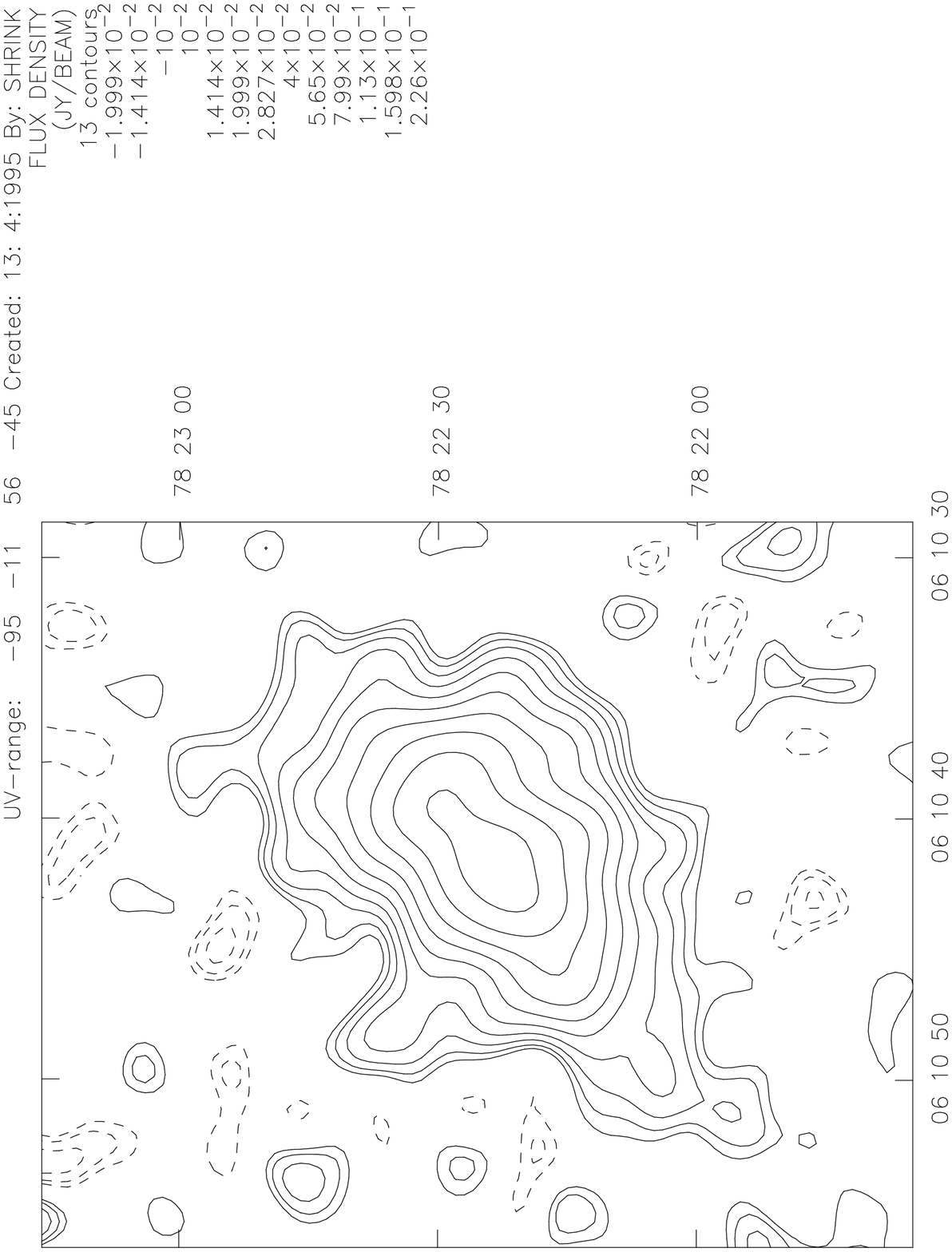,width=6.5cm,clip=,angle=270}{2}}
\caption[Map at 327 GHz]{
Map of NGC~2146 at 327 MHz at a resolution of 8 $\times$ 8 arcsec$^2$.
The contour levels are start at 10 mJy/beam, 
and increase by factors of $\sqrt 2$.
The  r.m.s. noise level is at 8 mJy/beam.
}
\end{figure}
\begin{figure}
\centerline{
\epsfigx{file=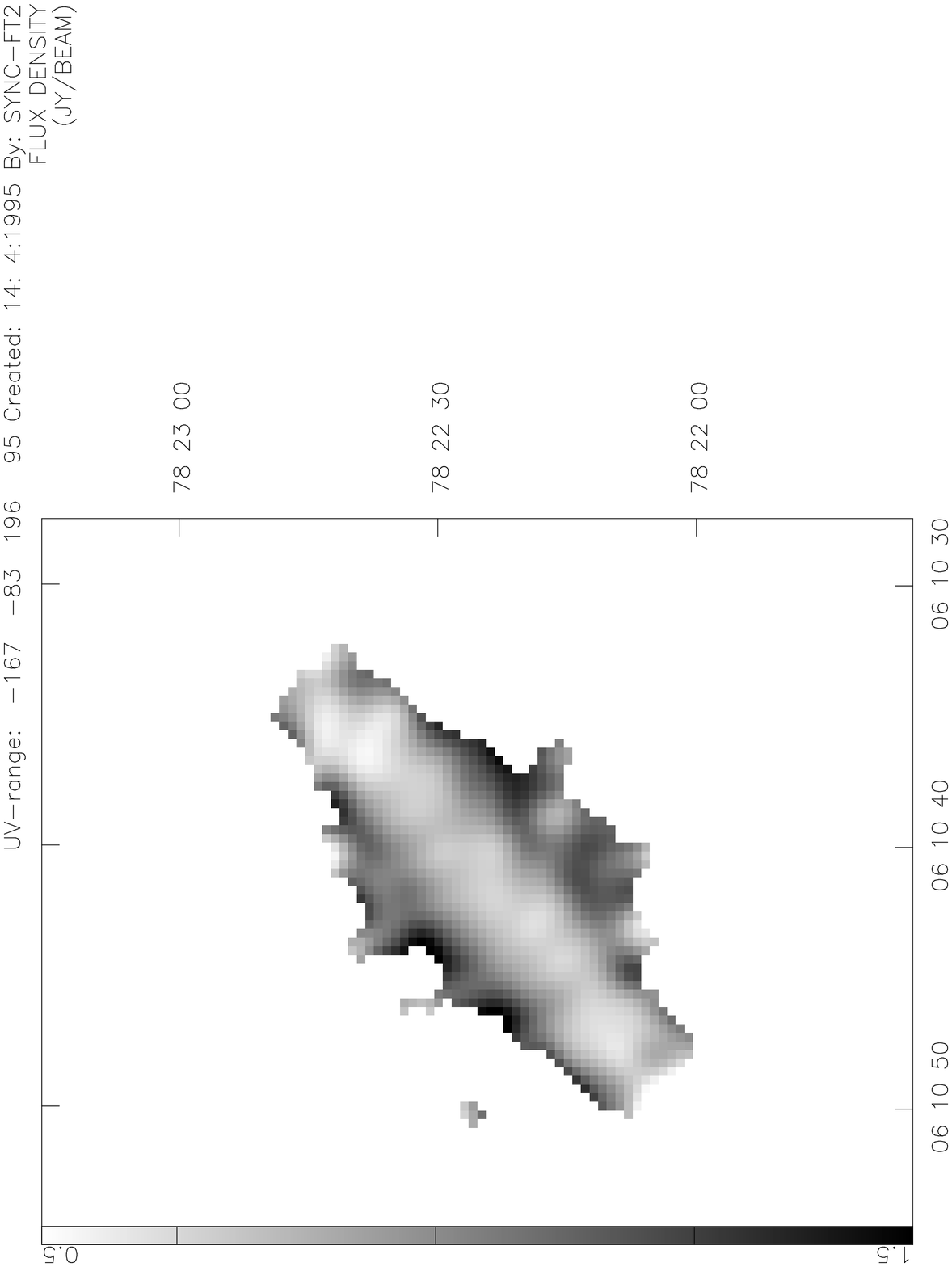,width=7cm,clip=,angle=270}{2}}
\caption[Spectral index map]{
Map of the spectral index of the radio emission between
1.5 and 8.4 GHz. The resolution is the same as in Fig. 1.
}
\end{figure}

\section{Modelling the radio emission}

Our aim in this work is to explain the multi-frequency
radio data by a combination of thermal emission and
non-thermal synchrotron radiation from CR electrons which have
propagated away from the sites where they have been accelerated.
The thermal component arises in {\HII} regions, whereas
supernova remnants (SNRs) are believed to be, at least from
energetic considerations, the most likely source
of CRs (e.g. Berezinsky et al. 1990, V\"olk 1992).

The propagation of the CR electrons
is described by a standard three dimensional diffusion
model (e.g. Ginzburg \& Syrovatskii 1964).
Other modes of propagation, such as convection, may also be important
especially in the halo (e.g. Hummel $\&$ Dettmar 1990)
at large distances from the galactic
plane, but for the sake of simplicity
we do not take this into account explicitly.  Our approach, therefore,
is to determine to what extent we can explain the observations using a
simple diffusion model alone.

The diffusion equation
describing the evolution of the electron distribution function
$f(E,\rvek,t)$ is as follows:

\begin{eqnarray}
&-&{\partial f(E,\rvek,t)\over \partial t}+
D(E)\bigtriangledown^2 f(E,\rvek,t) \nonumber\\
&+& {\partial \over\partial E}
\biggl( b(E) f(E,\rvek,t)\biggr) = - Q(E,t,\rvek)
\end{eqnarray}
where $D(E)$ is the diffusion
coefficient, assumed to be spatially homogeneous and isotropic.  The
diffusion coefficient is assumed to depend on the energy as
\begin{equation}
D(E)=D_0 \biggl( {E\over 1 {\rm GeV}} \biggr)^{\mu}
\end{equation}
which is in agreement with observations of CRs in our Galaxy (see e.g.
Berezinsky et al. 1990).
In the energy range relevant to our radio observations ($E>1$~GeV),
the energy losses of the electrons are
mainly synchrotron and inverse Compton
\begin{eqnarray}
b(E)&=&-{\diff E\over \diff t}\bigg|_{syn+IC}
={4\over 3} \sigma_{\rm T} c
\biggl({E\over m_ec^2}\biggr)^2(\urad+\ub) \nonumber \\
&\equiv& C\, E^2,
\end{eqnarray}
where $\sigma_{\rm T}$ is the Thompson scattering cross section,
$c$ the speed of light, $\ub$ the energy density of the magnetic field
and $\urad$ the energy density of
the radiation field below the Klein-Nishima limit (e.g. Longair 1992).

For the source function $Q(E,\rvek,t)$ we make the following assumptions:
\begin{enumerate}
\item The total source distribution can be separated and expressed
as the sum of individual sources i.e.
\begin{equation}
Q(E,\rvek,t)=\sum_i {\cal E}_i(E) {\cal R}_i(\rvek) {\cal T}_i(t).
\end{equation}
\item CRs are produced only by  SNRs from
massive stars ($m>5 \ms$). The source spectrum is
assumed to be a power-law in energy and to be identical for all sources
\begin{equation}
{\cal E}_i(E)=\biggl({E\over m_e c^2}\biggr)^{-\gamma}
\nu_{\rm SN} q_{\rm SN}.
\end{equation}
Here, $\nu_{\rm SN}$ is the SN rate and $q_{\rm SN}$ is the
number of relativistic electrons produced on average per energy interval
by a SNR.
\item The spatial distribution of a source is approximated
by an ellipsoid, gaussian distribution
\begin{eqnarray}
{\cal R}_i(\rvek)& &=\pi^{-{3\over 2}} (\rx \ry \rz)^{-{1\over 2}}
\nonumber \\
& &\exp\left\{-\left[{(x-x_i)^2\over \rx^2}+
{(y-y_i)^2\over \ry^2}+{(z-z_i)^2\over \rz^2} \right] \right\}
\end{eqnarray}
with ($x_i, y_i, z_i$) indicating the site of the centre of the source
and $\rx\ry\rz$ the axes of the ellipsoid.
In the following, we will identify the
major axes of NGC~2146 with the x-axis, the
direction within the galactic plane perpendicular to the
major axis with the y- and the direction perpendicular to the plane with
the z-axis.

\end{enumerate}

Eq.(1) can be solved for an infinite space
following the analysis of Syrovatskii (1959).
In a steady-state situation (i.e. for constant sources)
the electron distribution function is given by:
\begin{eqnarray}
f(E,\rvek)& =& \nu_{SN} q_{SN}
\sum_i \int^{E_2}_{max(E,E_1)} {\diff E_0\over C E^2}
\left({E_0\over m_e c^2}\right)^{-\gamma} \nonumber \\
 & &{\exp\left\{-\left[{(x-x_i)^2\over 4\lambda+ \rx^2}+
{(y-y_i)^2\over 4\lambda+ \ry^2}+
{(z-z_i)^2\over 4\lambda+ \rz^2} \right]\right\} \over
(4\lambda+\rx^2)^{1\over 2} (4\lambda+\ry^2)^{1\over 2}
(4\lambda+ \rz^2)^{1\over 2} }
\end{eqnarray}
where $E_2,E_1$ are the upper and lower end of the range of
electron injection energies  and $\lambda$ is given by:
\begin{eqnarray}
\lambda(E,E_0)&=&\int_E^{E_0} {D(E')\over b(E')} \diff E'
\nonumber \\
&=&{1\over C(1-\mu)} D_0
\left( E^{\mu-1}-E_0^{\mu-1} \right).
\end{eqnarray}

If temporal changes are important, a similar solution for
$f(E,\rvek,t)$ can be derived.
The synchrotron emission $\psyn(\nu,\rvek,t)$ is derived
from $f(E,\rvek,t)$ with the
simplifying assumption that each electron emits its synchrotron
radiation at the maximum frequency of the synchrotron spectrum.
\begin{equation}
\psyn(\nu,\rvek,t)=f(E(\nu),\rvek,t) {\diff E\over \diff t}
\bigg|_{syn} {\diff E\over \diff \nu}
\end{equation}
with the emitting frequency being
\begin{equation}
\nu= \left({E\over m_e c^2}\right)^2 \nu_G
\end{equation}
where $\nu_G = e B/m_e c^2$ is the gyrofrequency.

The thermal emission is also described as arising from
ellipsoidal sources, that are not necessarily
the same as the CR sources
\begin{equation}
\ptherm(\nu,\rvek) \propto \sum_i {\cal R'}_i(\rvek) \nu^{-0.1}.
\end{equation}
Both, $\psyn$ and $\ptherm$ have to be integrated through the halo along
the line of sight, taking into account the inclination angle of the galaxy,
and convolved with the beam.
The total radio emission, $\prad$, is finally obtained as
the sum of $\psyn$ and $\ptherm$.

\section{Fitting the model to the radio data}

In order to keep the diffusion model as simple as possible
we shall assume a steady state, i.e. we attempt to fit a model in which the
star formation rate and hence the sources of thermal and non-thermal
emission do not show any temporal variations.
In NGC~2146 the estimated life-time of
CR electrons is, due to the high radiation losses (see below), extremely
short (less than $2\times 10^6$ years). On these short time scales
the assumption of a steady state seems reasonable.
Furthermore we will assume as the
simplest case that {\HII} regions and supernova remnants
are co-spatial (at least at the resolution of our radio images).
This will be a good approximation if the sites of star formation have not
changed during the past $\approx 10^8$ years, i.e. the life-time
of supernova progenitors ($m \geq 5  \ms$).  If there has been
significant propagation of the star formation process through the
galaxy then this assumption must be modified.

Additionally we can constrain some of the parameters
directly from observational data.

1) The inclination of NCG~2146 is $\theta=65^\circ$
(Benvenuti et al. 1975).
Due to the disturbed appearance of NGC~2146 the inclination
angle is unlikely to be precisely determined; we shall adopt this
value and discuss this point more fully in Section 6.1.
The radius of the CR sources perpendicular
to the disk is taken to be $R_{z,i}=100$pc,
which is a typical scale height of the molecular gas disk (e.g.
Uns\"old \& Baschek 1991) in which we expect massive star formation
to be located.

2) We estimate the magnetic field
$B$ by applying a minimum energy calculation.
We adopt standard values for most parameters: a frequency range
of $10^7-10^{10}$ GHz;
radio spectral index 0.7
(here and in the following we define the spectral index
$\alpha$ as $\prad\propto \nu^{-\alpha}$);
ratio of total energy of CR's to total energy
of electrons $k=100$. For measured quantities we use
the emission at 1.5~GHz.  We obtain the volume of the
emitting region at this frequency by assuming that it
can be approximated by a cylinder with a height 1~kpc
and a radius of 45~arcsec ($\equiv$ 5kpc)
which corresponds to the lowest contour of the 1.5~GHz image.
With these assumptions we obtain $B=38 \mu G$.

\begin{figure*}
\begin{minipage}{150mm}
\centerline{
\epsfigx{file=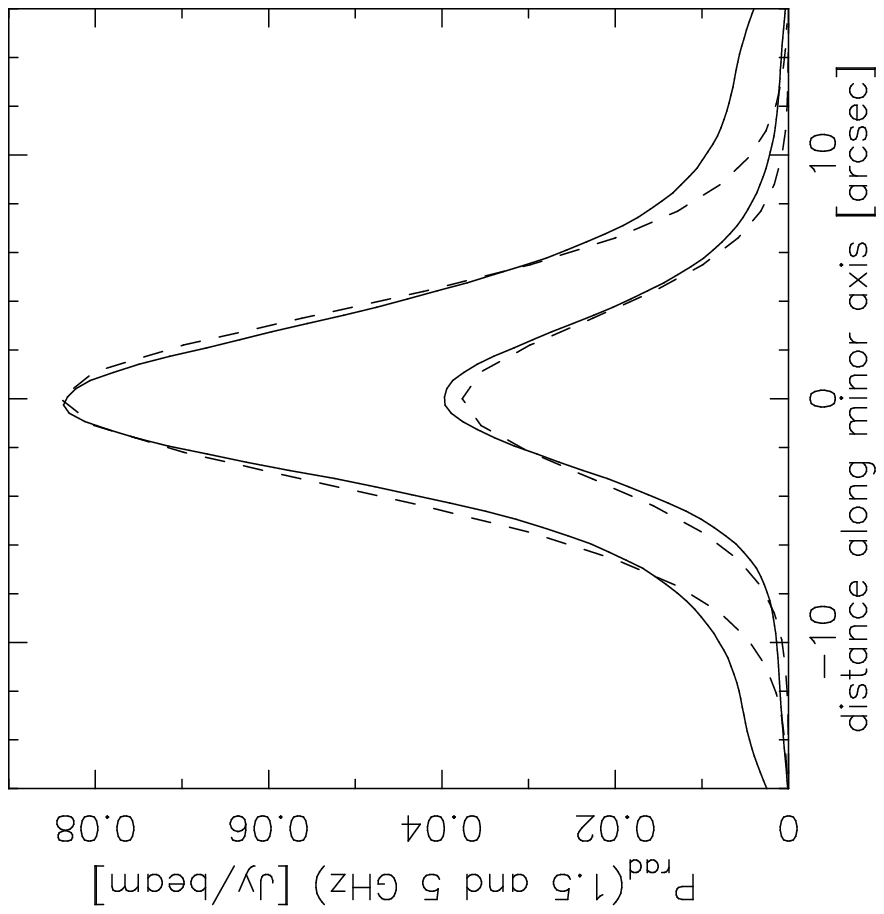,width=6.5cm,clip=,angle=270}{2}\quad
\epsfigx{file=fig_5b.ps,width=6.5cm,clip=,angle=270}{1}}
\smallskip
\centerline{
\epsfigx{file=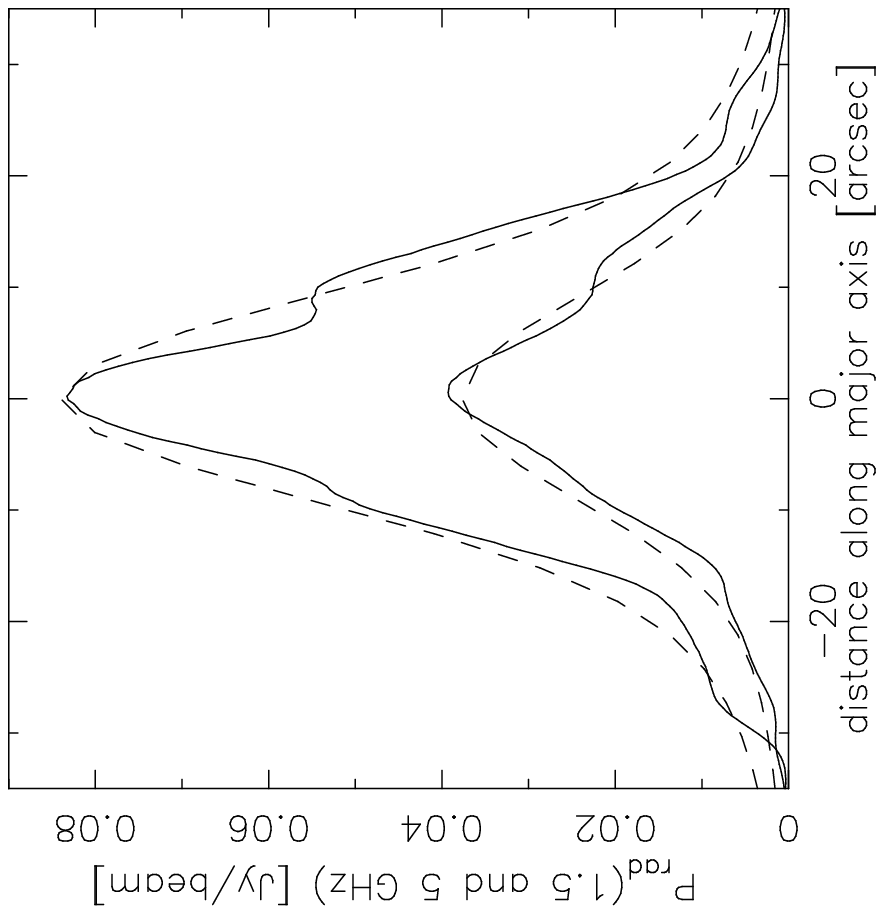,width=6.5cm,clip=,angle=270}{2}\quad
\epsfigx{file=fig_5d.ps,width=6.5cm,clip=,angle=270}{1}}
\caption[Model and data]{
Slices of the radio emission 
at  1.5, 5 and 8.4 and 15 GHz
along the minor and the major axis (full lines),
together with the best-fit steady state model results (dashed lines).
The parameters are:
$\gamma=1.7, \mu=0, D_0=2\times10^{28}$cm$^2$\,s$^{-1}$, 
$\ptherm(1.5 \rm GHz)=4$
mJy/beam at $\rvek=0$,
$R_{x,1}=1.3$ kpc, $R_{y,1}=0.7$ kpc,
$R_{x,2}=3.0$ kpc, $R_{y,2}=0.5$ kpc.
} 
\end{minipage}
\end{figure*}
The energy density of the radiation field, $\urad$, can be calculated from
the bolometric luminosity:
\begin{equation}
\urad={2 L_{bol} \over c\pi R^2}=
{\bigl(L_{bol}/ {\rm W}\bigr) \over \bigl(R/{\rm kpc}\bigr)^2}
\cdot 1.4 \cdot
   10^{-35}\biggl[{eV \over cm^3}\biggr].
\end{equation}
For NGC~2146 the greatest part of its luminosity is emitted in the
far-infrared (FIR);
we therefore approximate the bolometric luminosity $L_{bol}$ by
the far-infrared luminosity $\lfir$.
For the radius $R$ we take the radius derived from our radio data
($R=5$ kpc) as this
will correspond to the region in which most star formation is currently
occurring.  With these assumptions we obtain $\urad=20$ eV\,cm$^{-3}$.
The energy density of the radiation field, $\urad$, can be calculated from
the bolometric luminosity:
\begin{equation}
\urad={2 L_{bol} \over c\pi R^2}=
{\bigl(L_{bol}/ {\rm W}\bigr) \over \bigl(R/{\rm kpc}\bigr)^2}
\cdot 1.4 \cdot
   10^{-35}\biggl[{eV \over cm^3}\biggr].
\end{equation}
For NGC~2146 the greatest part of its luminosity is emitted in the
far-infrared (FIR);
we therefore approximate the bolometric luminosity $L_{bol}$ by
the far-infrared luminosity $\lfir$.
For the radius $R$ we take the radius derived from our radio data
($R=5$ kpc) as this
will correspond to the region in which most star formation is currently
occurring.  With these assumptions we obtain $\urad=20$ eV\,cm$^{-3}$.

With these derived values,
the fitting of the diffusion model now proceeds by fitting to
slices extracted from our multi-frequency radio images
along both the major and minor axis of NGC~2146.
In Figs. 5 and 6 we show the observational data
and the corresponding model fits.

In constraining the parameters for the model we note some
general features of the data.

i) The spectral index between 1.5 and 8.4 GHz is approximately constant
along the major axis (Fig. 8a); this suggests that
the CR source is itself extended in this direction unless
the diffusion coefficient had a strong energy dependence with
$\mu \approx 1$. (In this latter case
$\lambda$ (Eq. 8) would be largely independent of the electron energy
and hence the spatial distribution of the electrons
would also be energy independent (see Eq. 7)).

ii) The steepening of the spectrum away from the major axis (Fig. 8)
indicates that $\mu$ is considerably less than one and that the
electrons must have suffered from energy losses.

iii) The radio flux decreases away from the centre of the galaxy both
along the major and minor axis (Figs. 5 and 6).

Taken together, these observations
suggest that the model should have the following
general characteristics:
\begin{itemize}
\item The source distribution is extended along the major axis but not
along the minor axis, i.e. it has the shape of a bar.
\item The energy dependence of the diffusion coefficient, $\mu$, must be
considerably less than unity, and is constrained by the data along the minor
axis of NGC~2146.
\end{itemize}

\begin{figure*}
\begin{minipage}{150mm}
\centerline{
\epsfigx{file=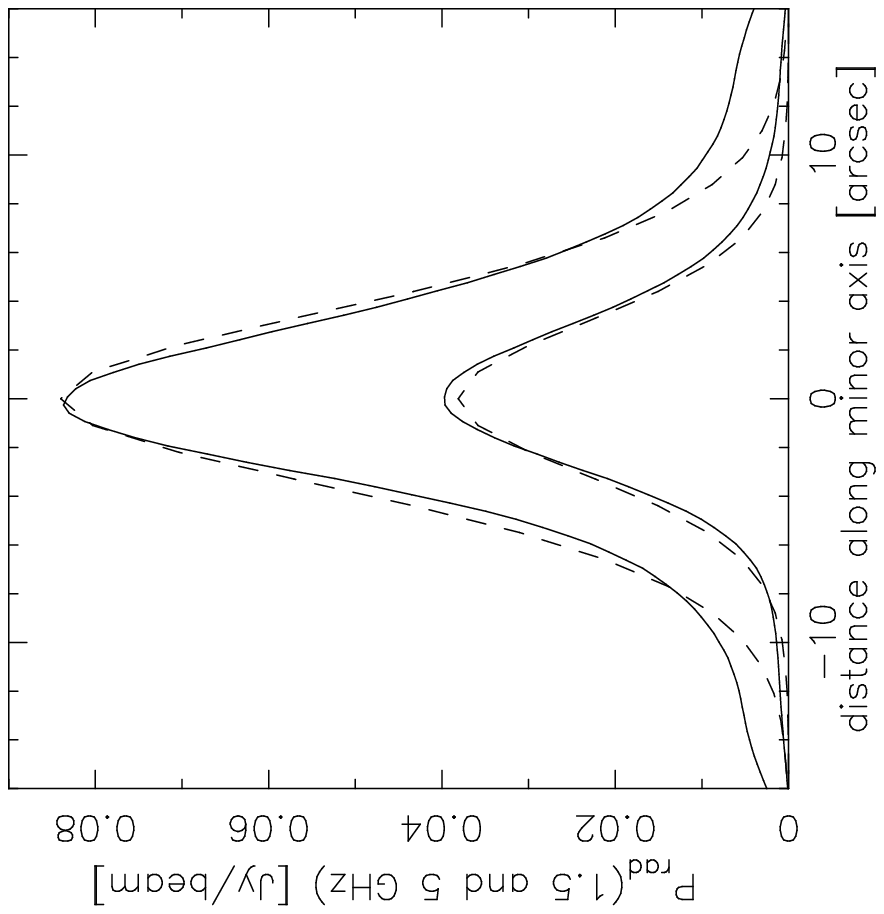,width=6.5cm,clip=,angle=270}{2}
\epsfigx{file=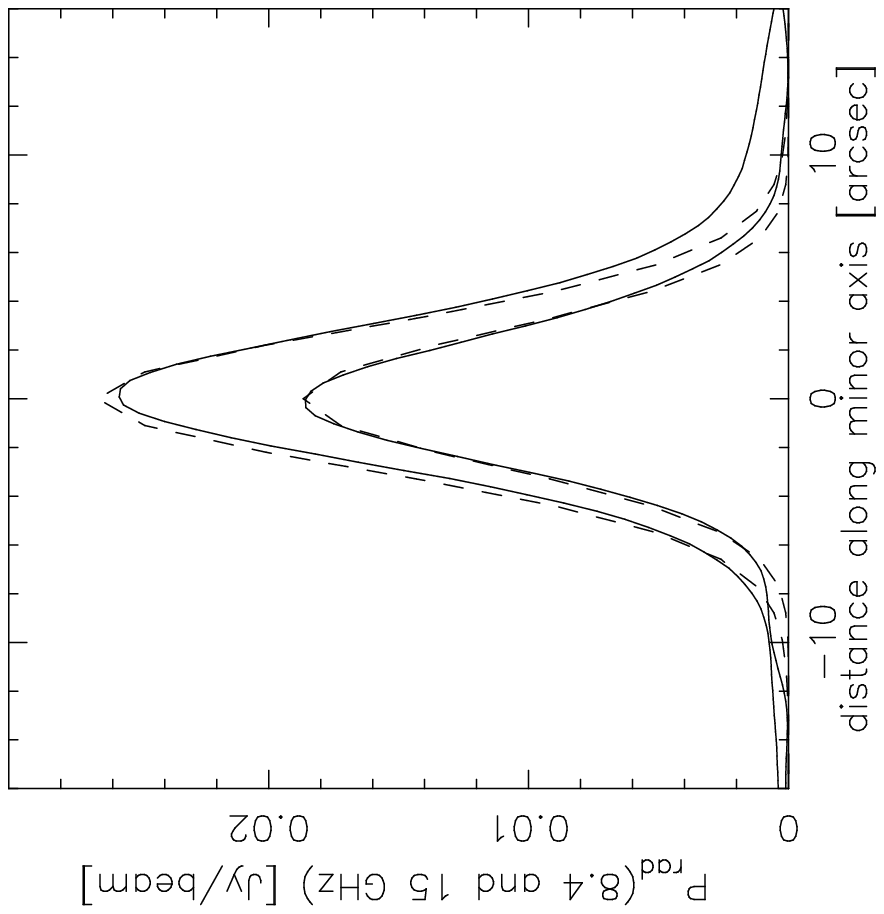,width=6.5cm,clip=,angle=270}{2}}
\smallskip
\centerline{
\epsfigx{file=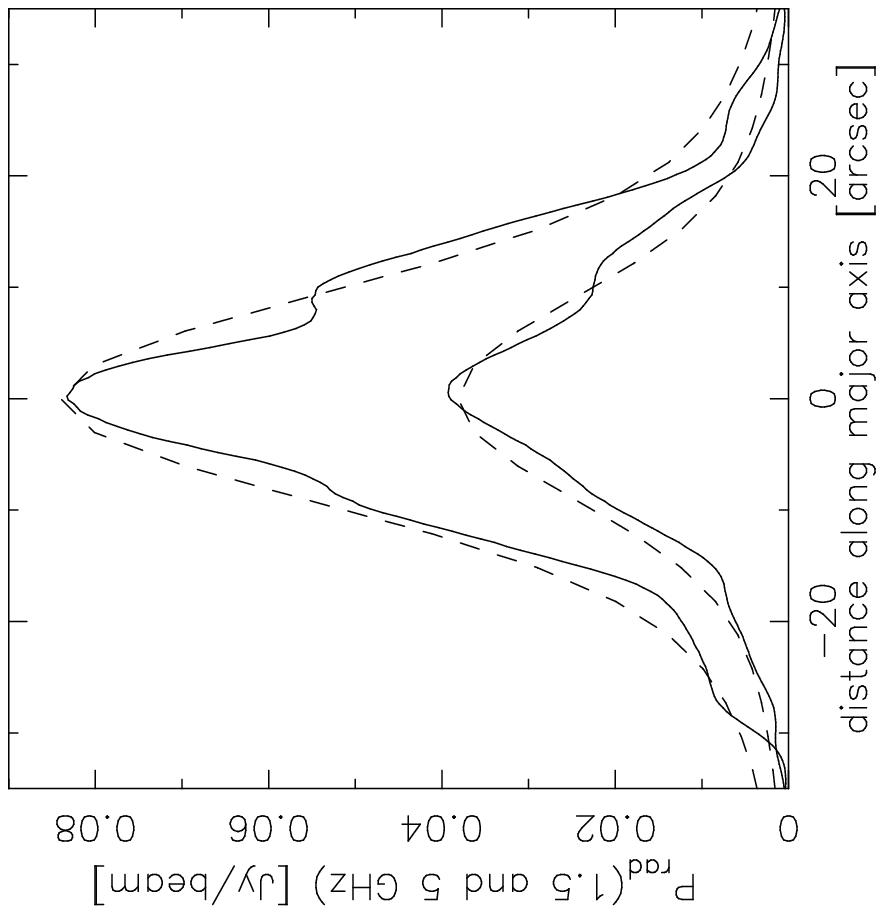,width=6.5cm,clip=,angle=270}{2}
\epsfigx{file=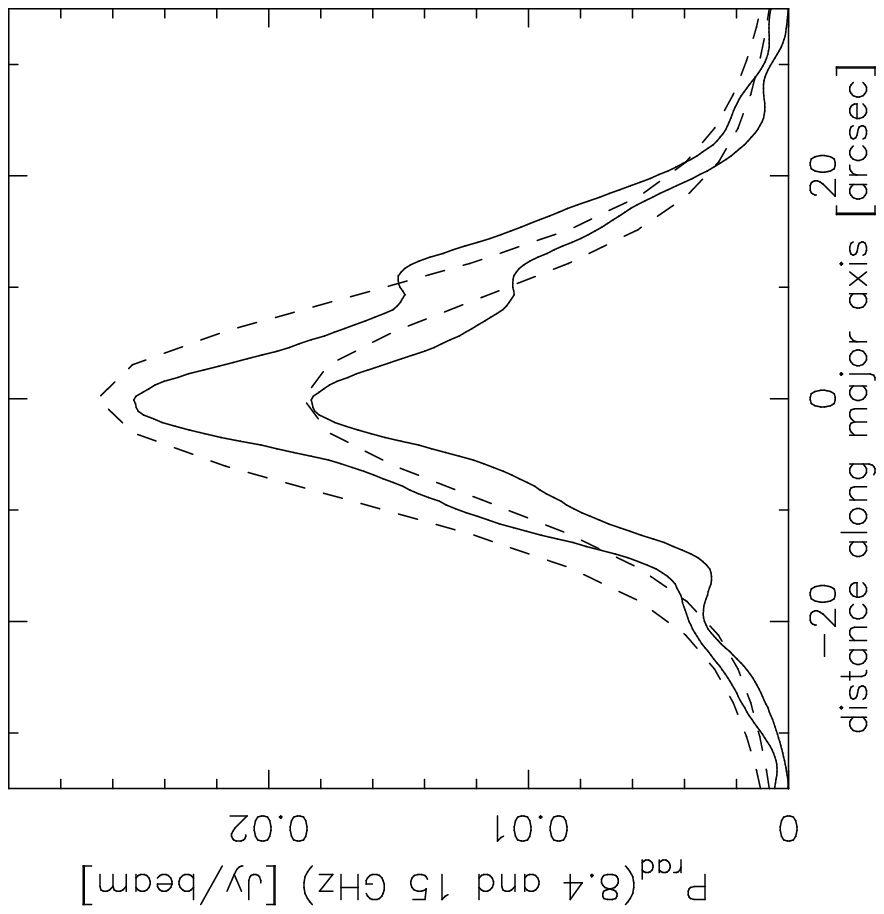,width=6.57cm,clip=,angle=270}{2}}
\caption[Model and data]{
Slices of the radio emission (full lines),
together with the best-fit results 
for a model with temporal changes in the injection rate (dashed lines).
The parameters are: 
$\gamma=1.65,
\mu=0, D_0=2\times10^{28}$cm$^2$\,s$^{-1}$, 
$\ptherm(1.5 \rm GHz)=23$
mJy/beam at $\rvek=0$, the source structure is the same as in Fig. 5.
The injection of CR's has stopped $5.7\times10^5$ years ago.
} 
\end{minipage}
\end{figure*}
The fitting of the model to the observational data now proceeds by
a straightforward search of parameter space for model parameters which
lead to the best fit of the multi-frequency radio data.  The model
most strongly constrains the spatial distribution and strength of the
thermal and non-thermal emission regions, and also the energy dependence
of the diffusion coefficient.  Our results and conclusions for the
simplest form of the model without temporal variations are shown in
Fig.~5 and can be summarized as follows.

1) The source structure cannot be adequately
represented by a single ellipsoidal emission region. Two sources,
both centred at the maximum of the radio emission,
are necessary to account for the data.
Our best fit parameters for these two sources are:
\begin{description}
\item[-] source 1: $R_{x,1}=1.3$ kpc, $R_{y,1}=0.7$ kpc, relative intensity 1
\item[-] source 2: $R_{x,2}=3.0$ kpc, $R_{y,2}=0.5$ kpc, 
relative intensity 0.3
\end{description}
We interpret this  result as suggesting that the distribution of
CR sources and hence star formation activity is in the form of a bar.
The parameters describing source 1 are relatively well constrained
by the shape of the radio emission (for a fixed value of the
galaxy inclination).
The parameters of source 2 are less well constrained, however this
has a relatively minor effect on the other model parameters.

\begin{figure*}
\begin{minipage}{150mm}
\centerline{
\epsfigx{file=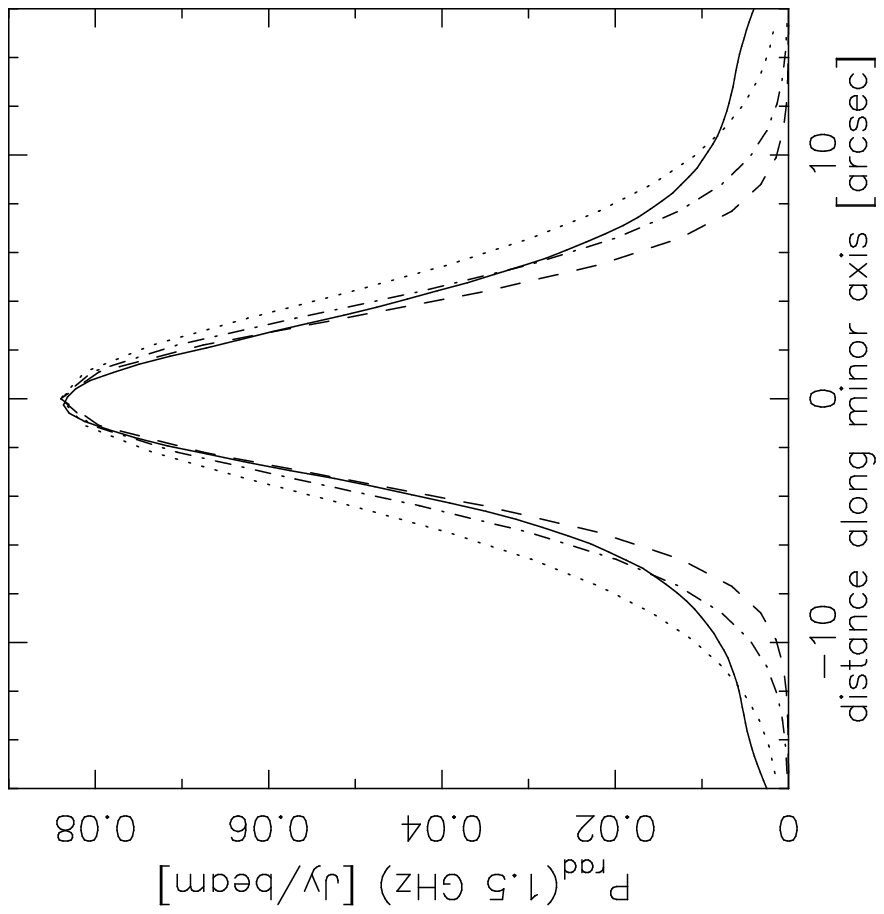,width=6.5cm,clip=,angle=270}{2}\quad
\epsfigx{file=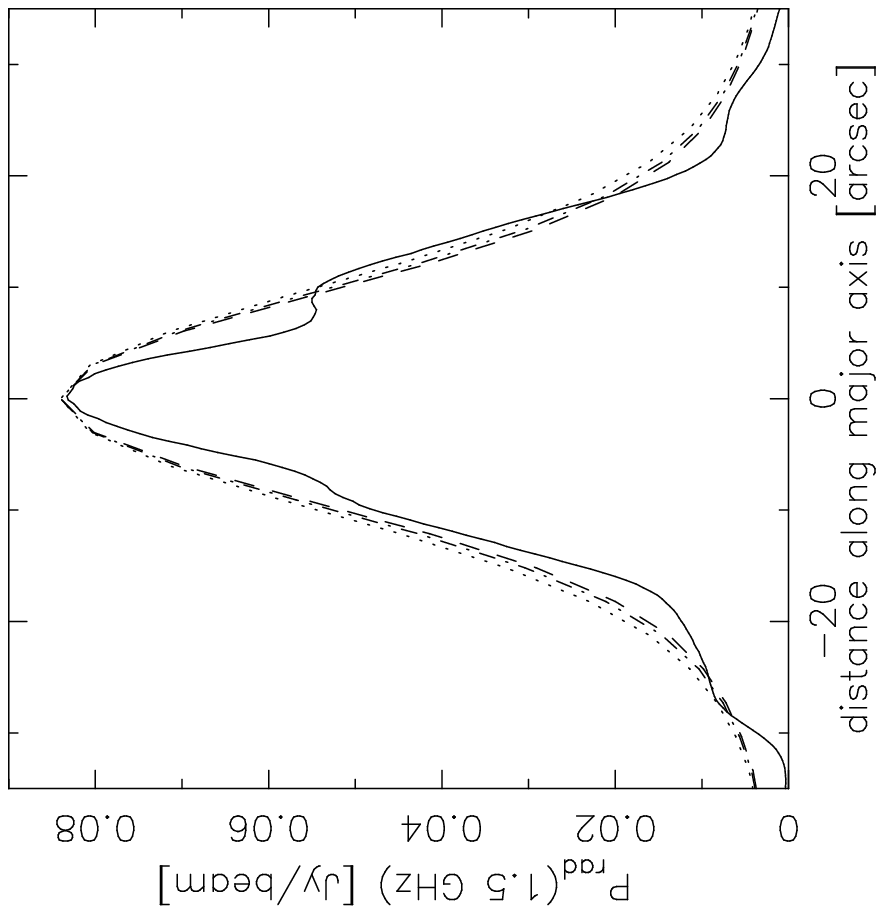,width=6.5cm,clip=,angle=270}{2}}
\caption[Diffusion coefficient]{
Slices of the radio emission at 1.5 GHz 
along the minor and 
the major axis (full line)
together with model results for different values
of the diffusion coefficient:
$D_0=1.0\times 10^{28}$cm$^2$\,s$^{-1}$ (dashed line),
$D_0=2.0\times 10^{28}$cm$^2$\,s$^{-1}$ (dashed-dotted line),
and $D_0=4.0\times 10^{28}$cm$^2$\,s$^{-1}$(dotted line).
The other parameters are as in Fig. 5.
}
\end{minipage}
\end{figure*}
\begin{figure*}
\begin{minipage}{150mm}
\centerline{
\epsfigx{file=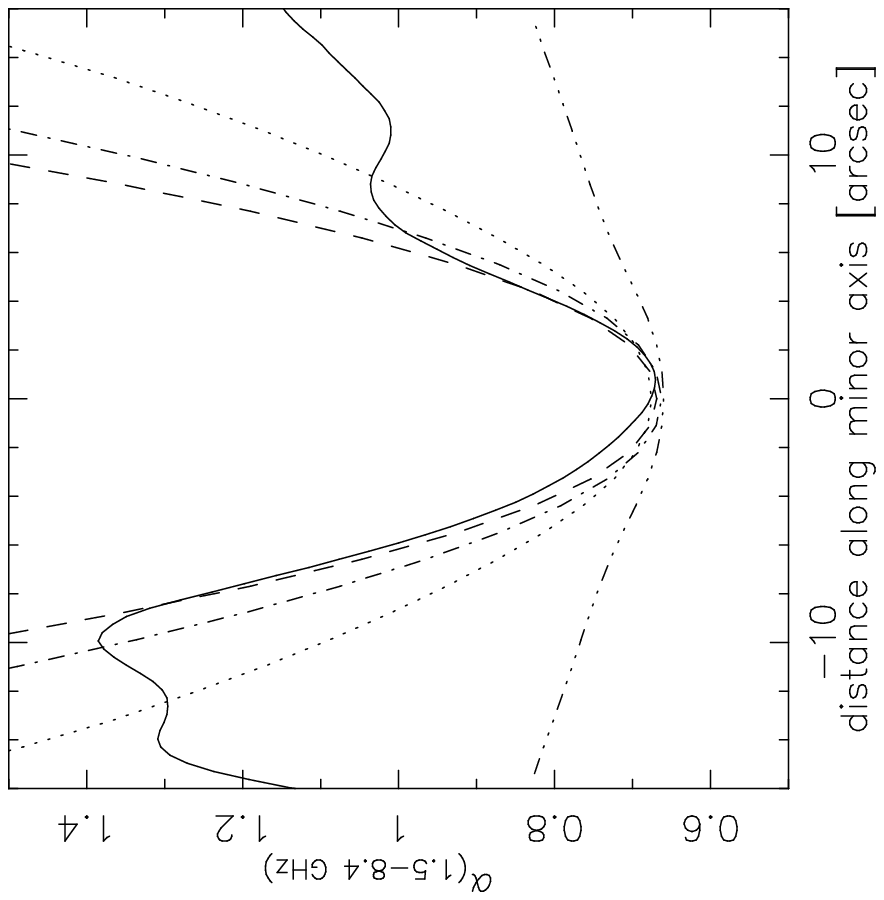,width=6.5cm,clip=,angle=270}{2}\quad
\epsfigx{file=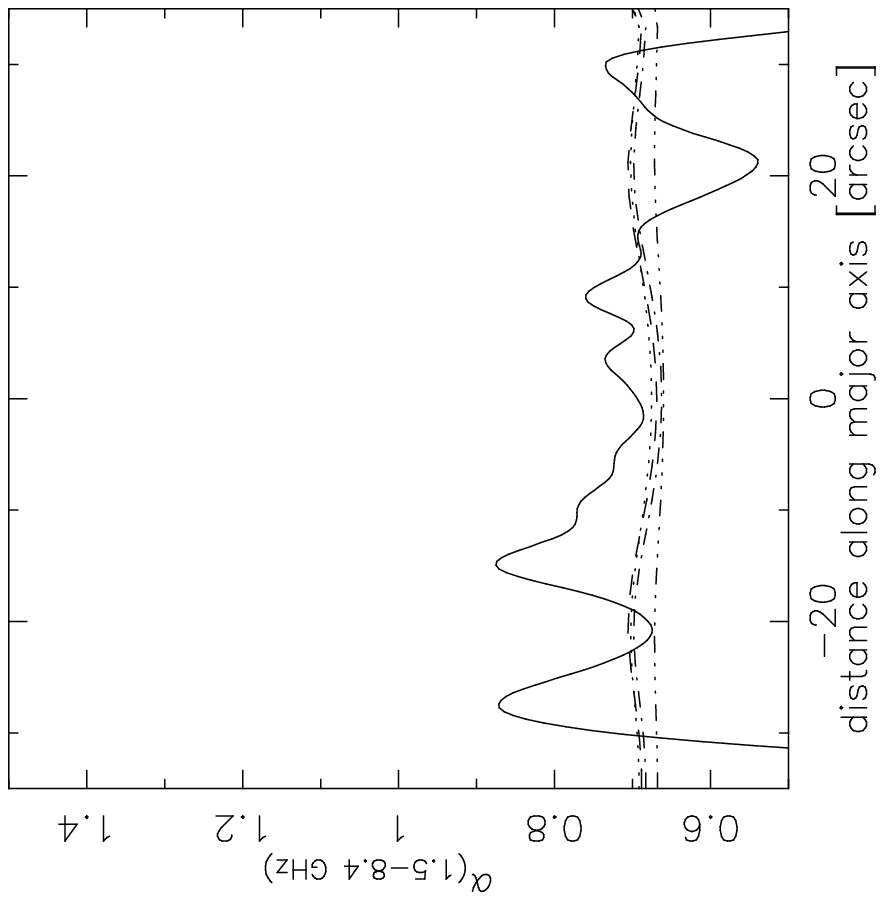,width=6.5cm,clip=,angle=270}{2}}
\caption[Engery dependence]{
Slices along the minor and the major axis 
of the spectral index  between
1.5 and 8.4 GHz, $\alpha_{(1.5-8.4GHz)}$ (full line),
together with the model results for different values of $\mu$.
The parameters are:
$\mu=0, \gamma =1.7, D_0=2.0\times 10^{28}$cm$^2$\,s$^{-1}$ (dashed line),
$\mu=0.2, \gamma=1.65, D_0=1.5\times 10^{28}$cm$^2$\,s$^{-1}$ (dashed-dotted line),
$\mu=0.5, \gamma=1.6, D_0=0.8\times 10^{28}$cm$^2$\,s$^{-1}$ 
(dashed-dotted line),
and $\mu=0.9, \gamma=1.45, D_0=0.3\times 10^{28}$cm$^2$\,s$^{-1}$
(three dots - dash).
The values of $\gamma$ and $D_0$ have to be altered in order to
ensure that the fits to the radio emission from Fig. 5 are
identical for the different values of $\mu$.
The source distribution is parameterized as in Fig. 5.
}
\end{minipage}
\end{figure*}

2) The injection index of the electron population is $\gamma=1.7$ and
we infer a low thermal fraction of $\ptherm/\prad =17 $ per cent at 15 GHz.
The thermal fraction is constrained by the radio spectrum at the
centre of the galaxy.

3) It is clear from Fig. 5 that we are not able,
with the simplest form of the model, to get good
fits to the data over the whole frequency range: At 5, 8.4 and 15~GHz
the model predictions for the peak radio emission are either 
slightly too low or too high. 
The discrepancies however never exceed 10 $\%$ and
it should furthermore be noted that the {\it shape}
of the radio emission can be reproduced well at all frequencies.

The results obtained by allowing for temporal variations in the
star formation rate, while keeping the
source distribution constant are shown in Fig. 6.
In this case, an injection index
of $\gamma=1.65$ is used and the temporal variations are assumed to
have a simple form with
injection of CR electrons having stopped
$5.7\times 10^5$ years ago.  The predicted thermal contribution
is now much greater than in the steady-state model.  At 15~GHz the emission
is almost completely thermal and at 1.5 GHz the thermal
fraction is still $P_{therm}/P_{rad}=27$ per cent.
The time-dependent solution produces an
improved model fit as it permits an aged electron population
to develop.
In general, for both the steady-state and time-dependent cases,
the fits along the major axis are worse than those along
the minor axis --- this is because of the structure of
$\prad$ which reflects the complicated intrinsic source structure
(see Fig. 2) which we have not taken into account.  However,
this does not significantly effect are conclusions concerning propagation
of CR to be discussed below.

4) The value of the diffusion coefficient $D_0$, and its energy
dependence, is strongly constrained by the data perpendicular to the
inferred bar within NGC~2146 along the minor axis.
Furthermore, the results are
very similar for both the steady-state and a time-dependent models.
In Fig. 7 we show the results for the steady-state case -- the best
fit for the diffusion coefficient is
$D_0=1.0 \rightarrow 4.0 \times 10^{28} $cm$^2$\,s$^{-1}$, with $\mu=0$.
The shape of the spectral index distribution along the minor axis depends
sensitively on $\mu$, which is illustrated in Fig. 8.
Our best estimate for $\mu$ is in the range $\mu=0.0 \rightarrow 0.2$
with a firm upper limit of $\mu < 0.5$.

\section{Discussion}

\subsection{Model uncertainties}

As illustrated in Fig. 5-8 the model describes the data
well and allows us to constrain important parameters such as the
diffusion coefficient and its energy dependence 
and infer the existence of a central bar.
The error in the determination of these parameters
depends however not only on the quality of the model fit, but
also on a number of parameters that were
determined independently such as $B, \urad$, the inclination angle and distance
of the NGC~2146.

In order to fit the spectral data a model that produces a break in the
synchrotron spectrum is required -- this is evident from the data since
the spectrum at the centre of NGC~2146
steepens between 5 and 8.4 GHz before flattening towards 15 GHz which we
attribute to thermal emission.
The simplest way to achieve such a break is to assume some temporal dependence
of the CR injection and hence of the star-formation rate.
The time-scale of the changes implied by the data
is rather short, less than $10^6$ years, however there is a strong
dependence on $B$ and $\urad$: $t\propto B^{1/2}/(\urad+\ub)$.
Whereas our estimate for $\urad$ does not suffer from large 
uncertainties, $B$ is
very uncertain and an improved estimate would require an independent
determination of the magnetic field.

The main sources of error for the diffusion coefficient,
$D_0$, are the uncertainty
in $\urad$, $B$, the distance to NGC~2146 and the assumed structure of
the emitting region.
In principle it could be possible to increase/decrease the size
of the CR source along the minor axis ($R_{y}$) and
try to fit the data by decreasing/increasing $D_0$.
If $R_y$ is very small, even with a large value of $D_0$ the model
cannot produce the observed shape of $\prad$.
On the other hand, a larger assumed source size would make it
impossible to fit the spectral variations unless we assumed that either
the injection index of the CR population varied within the disc of the
galaxy, and/or that different regions of the galaxy 
had undergone very different
star-formation histories.  Within the range of acceptable $R_{y}$ the
estimated values for $D_0$ all lie within our quoted error bounds.
The uncertainty in the inclination angle of NGC~2146
does not affect our estimates of $D_0$ and $\mu$ greatly and  it 
most strongly affects the assumed source size. 
Our conclusion regarding the bar-like shape of
the star forming region  depends therefore sensitively on
the assumed inclination angle.
However, only for a severe underestimate of the true inclination
angle of NGC~2146 (i.e. if $i\ga 80^\circ$) this conclusion cannot
be maintained. In this case the radio emission along
the minor axis would be almost entirely from the halo.

\subsection{CR propagation in the central disk and halo}

The diffusion model is a good fit to the radio data for NGC~2146
along the major axis of the galaxy and along the minor axis for distances
less than 10 arcsec (2.6~kpc) from the major axis.
This suggests that diffusion
is the dominant mode of propagation in this region which corresponds to
the inner, most actively star forming disc of NGC~2146.

At distances greater than 10 arcsec from the centre of the galaxy
along the minor axis the model is no longer an adequate fit to the data
(Figs. 5 and 6).
The shape of the 1.5~GHz emissivity is very flat
suggesting that transport of CR electrons is more efficient than
would be expected from diffusion alone.
Furthermore, the spectral index flattens which indicates that (i) the
energy losses have decreased rapidly, or (ii) the diffusion coefficient
increases, or (iii) a much faster process than diffusion is responsible
for the transport of the CR electrons.
Upon examination of the radio emission in the region where the
diffusion model begins to fail we note that the radio images
(Figs. 1 and 3) show filamentary structure.
The most likely reason for this behaviour is that
away from the centre of the galaxy a large fraction of the radio
emission is from a halo which is seen in projection.

The propagation of CR electrons in the halo might be  more
complicated than we have considered for the propagation within the disc.
Evidence for this comes from observations of the
edge-on galaxies NGC~891 and NGC~4631 where extended radio
continuum emission in the halo with a
rather flat spectral index
has been observed (Hummel 1991). Breitschwerdt (1994) has
explained this flat spectral index in a
galactic wind model including diffusive and convective transport of
CR's. Siemieniec and Cesarsky (1991) modelled the spectral
index in the halo of NGC 891
by a diffusion model with an
outwardly increasing diffusion coefficient.
Both galaxies show that in the halo the propagation of CR electrons
is faster than that predicted by steady-state diffusion alone.

Due to the strong star formation activity in NGC~2146,
an outflow from the disc triggered by
correlated SN explosions  may be present,
a mechanism which
is variously described as a  ``galactic fountain'' (Shapiro \& Field 1976)
or ``chimney'' model (Ikeuchi 1988, Norman \& Ikeuchi 1989).
Such an outflow would transport CR's and magnetic field into the
halo and
could therefore account for a higher radio emission in the halo
than expected from a pure diffusion model.
The analysis of X-ray and optical data of NGC~2146 (Armus et al. 1995)
has shown that there are indeed
indications of the existence of such a starburst-driven superwind.

\subsection{Star formation in NGC~2146}

Hutchings et al. (1990) proposed that NGC~2146 is in the late stage of
a merger. The merger must have begun $\geq 10^9$ years ago
as the rotation curve does not appear to be disrupted in the outer
regions of the galaxy (Young et al. 1988a, show the inner regions
of the galaxy to have non-circular motions of the order of 100 km s$^{-1}$).
In this scenario material has collapsed into the centre of the galaxy
and a starburst has commenced in the nucleus.

The data and model
presented in this paper further emphasize that active star formation is
confined only to the centre of the galaxy. The radio images show
that the star formation activity is located in a
peculiar spiral {\bf S} shape around the peak of the radio emission.
The merger may have triggered the formation of
a bar which feeds the nuclear starburst with material and is
responsible for the unusual structure of the radio emission. Star formation
is proceeding all along the central bar, and the nucleus
is not prominent in the high-resolution radio images.
Young et al. (1988a) suggest that if the galaxy is in the
late stage of a merger the  nucleus of the merging companion could
be obscured by the dust lane. The images of the radio emission
presented here argue against this as we do not find two spatially distinct
peaks of radio emission  which could be identified as the
nuclei of the merging systems. Instead, the star formation is
distributed throughout the bar-like central region.  The unresolved sources
seen on our highest resolution images have sizes of less than 100~pc and
may represented isolated pockets of star formation or could be individual
high-luminosity supernovae as seen in M82 (Muxlow {\em et al.} 1994).
The presence of dynamical features such as the arm in the H$\alpha$
emission (Young, Kleinmann \& Allen 1988b)
and the bar in the radio emission suggest that this starburst
is dynamically driven, i.e. due to a merger or an interaction.

{\bf Acknowledgements:}
The VLA is operated by the National Radio Astronomy Observatory for
Associated Universities Inc., under a cooperative agreement with the
National Science Foundation. UL gratefully acknowledges the receipt of
a postdoctoral fellowship of the
Deutsche Forschungsgemeinschaft (DFG).
We would like to thank the referee,
Prof. Davies, for useful comments.


\begin{thebibliography}{}

\bibitem[]{armus}
Armus L., Heckman T.M., Weaver K.A., Lehnert M.D., 1995, \apj 445,666
\bibitem[]{benv} 
Benvenuti P., Capacioli M., D'Odorico S., 1975, \aajou 41, 91
\bibitem[]{bere}Berezinsky V.S., Bulanov S.V., Ginzburg V.L.,
Dogiel V.A., Ptuskin V.S., 1990, {\it Astropysics of Cosmic Rays},
North Holland
\bibitem[]{breit}
Breitschwerdt D., 1994, Habilitationsschrift, Universit\"at Heidelberg
\bibitem[]{condon} Condon J.J., 1983, \apjs 53, 459
\bibitem[]{ginz} 
Ginzburg V.L., Syrovatskii S.I., 1964, {\it The Origin of Cosmic Rays},
Pergamon Press, Oxford
\bibitem[]{hutch} 
Hutchings J.B., Neff S.G., Stanford S.G., Lo E., Unger S.W.,
1990, \aj 100, 60
\bibitem[]{hummel}
Hummel E., 1991 in {\it The Interstellar Disk-Halo Connection in
Galaxies}, ed. H. Bloemen, Kluwer, 257
\bibitem[]{humdett} Hummel E., Dettmar R.-J., 1990, \aajou 236, 33
\bibitem[]{ikeuchi} Ikeuchi S., 1988, Fund. Cosmic Phys., 12, 255
\bibitem[]{kronberg}
Kronberg P.P., Biermann P., 1981, \apj 243, 89
\bibitem[]{lerche}
Longair M.S., 1992, {\it High Energy Astrophysics}, Vol. 1,
Cambridge University Press
\bibitem[]{mux}
Muxlow T.W.B., Pedlar A., Wilkinson P.N., Axon D.J.,
Sanders E.M., DeBruyn A.G., 1994, \mnras 226, 455
\bibitem[]{norman}
Norman C. A., Ikeuchi A., 1989, \apj 345 372
\bibitem[]{pohl}
Siemieniec G., Cesarsky C., 1991, \aajou 245, 418
\bibitem[]{shapiro}
Shapiro P.R., Field G.B., 1976, \apj 205, 762
\bibitem[]{syro}
Syrovatskii S.I., 1959, Soviet Astronomy Vol. 3, No. 1., 22
\bibitem[]{volk}
V\"olk H.J., 1992, in {\it Particle Acceleration in
Cosmic Plas\-ma},
 eds. G.P. Zank and T.K. Gaisser, AiP Conf. Proc. 264, AiP, New York, 199
\bibitem[]{unsold}
Uns\"old A.,Baschek B., 1991, {\it The New Cosmos}, Springer-Verlag,
Heidelberg
\bibitem[]{youngb}
Young J.S., Kleinmann S.G., Allen L.E., 1988b, \apj 334, L63
\bibitem[]{younga}
Young J.S., Claussen M.J., Kleinmann S.G., Rubin V.C., Scoville N.,
1988a, \apj 331, L81

\end{thebibliography}
\end{document}